\documentstyle[12pt,aasms4]{article} 

\def\la{\lower.5ex\hbox\frac{$\; \buildrel <}{\sim \;$}}
\def\ga{\lower.5ex\hbox\frac{$\; \buildrel >}{\sim \;$}}

\begin{document}

\title{THREE-DIMENSIONAL ORBITS OF METAL-POOR HALO STARS AND
 THE FORMATION OF THE GALAXY}

\author{Masashi Chiba}
\affil{National Astronomical Observatory, Mitaka, Tokyo 181, Japan}

\and

\author{Yuzuru Yoshii\altaffilmark{1}}
\affil{Institute of Astronomy, Faculty of Science, University of Tokyo, 
Mitaka, Tokyo 181, Japan}
\altaffiltext{1}{Also at Research Center for the Early Universe,
Faculty of Science, University of Tokyo, Bunkyo-ku, Tokyo 113, Japan}

\begin{abstract}
We present the three-dimensional orbital motions of metal-poor stars
in conjunction with their metal abundances, for the purpose of getting
insight into the formation process of the Galaxy. Our sample stars, which
include metal-deficient red giants and RR Lyrae variables observed by
the Hipparcos satellite, are least affected by known systematics,
stemmed from kinematic bias, metallicity calibration, and secondary metal
contamination of stellar surface.
We find, for the stars in the metallicity range of [Fe/H]$\le -1$, that
there is no evidence for the correlation between [Fe/H] and their orbital
eccentricities $e$.
Even for [Fe/H]$\le -1.6$, about 16\% of the stars have $e$ less than 0.4.
We show that the $e$ distribution of orbits for [Fe/H]$\le -1.6$
is independent of the height $|z|$ away from the Galactic plane, whereas
for [Fe/H]$>-1.6$ the stars at $|z|\ge 1$ kpc are systematically devoid of
low-$e$ orbits with $e \la 0.6$.
This indicates that low-$e$ stars with [Fe/H]$\le -1.6$ belong to the halo
component, whereas the rapidly-rotating thick disk with a scale height
$\sim 1$ kpc has a metal-weak tail in the range of $-1.6<$[Fe/H]$\le -1$.
The fraction of this metal-weak thick disk appears to be only
less than 20\%.
The significance of these results for the early evolution of the Galaxy
is briefly discussed.
\end{abstract}

\keywords{stars: Population II --- Galaxy: formation --- Galaxy: halo
--- Galaxy: kinematics and dynamics}

\section{INTRODUCTION}

The most useful diagnostics for probing how the Galaxy has formed
lie in the relationship between metal abundances [Fe/H] and orbital
eccentricities $e$ of stars (e.g. Freeman 1987; Majewski 1993).
For example, if the Galaxy underwent a monolithic, rapid collapse,
the orbits of stars which have formed from free-falling gas clouds were made
highly eccentric in the course of a rapid change of the Galactic mass
concentration. The aftermath of collapse therefore prohibits the existence of
metal-poor stars having nearly circular orbits.
On the other hand, if the mass concentration occurred slowly in a manner of
either a quasi-static contraction or merging/accretion of numerous fragments,
some fraction of metal-poor stars are inevitably expected to hold nearly
circular orbits, where the eccentricities have been approximately preserved
up to the present epoch.

However, our knowledge of the Galactic past remains still far from reaching
general consensus since the pioneering work by Eggen, Lynden-Bell \& Sandage
(1962, hereafter referred to as ELS).
ELS showed, based on the kinematically selected sample of stars,
the evidence supporting the picture of a rapid Galaxy collapse
from their finding of a strong correlation between [Fe/H] and $e$.
In contrast, Yoshii \& Saio (1979) and Norris, Bessell \& Pickles (1985, NBP)
demonstrated the existence of low-$e$, low-[Fe/H] stars, and argued that
the ELS result is an artifact of selecting metal-poor stars having high
eccentricities.
This divergence of the results stems from how well the existing systematics
are controlled when selecting the samples for the analysis.
In particular, the systematic biases inherent in the samples either create
or diminish a correlation between [Fe/H] and $e$ of stars in a non-trivial
manner that the picture for the initial contraction of the Galaxy is
drastically changed.

It has recently been claimed that a non-negligible fraction of low-$e$,
low-[Fe/H] red giants obtained by NBP are dropped out because of their
underestimated calibration of [Fe/H] (Twarog \& Anthony-Twarog 1994;
Ryan \& Lambert 1995). 
In Figure~1(a), we reproduce their result on the relation between [Fe/H] and
planar $e$ projected onto the Galactic plane, for dwarfs (crosses) and
red giants (filled symbols),
where the revised [Fe/H] for the latter stars are used if they are available
from either Anthony-Twarog \& Twarog (1994, ATT) or Ryan \& Lambert (1995).
It is suggested that the revision for [Fe/H] of some red giants turns out to
decrease a number of low-$e$, low-[Fe/H] stars, thereby weakens the conclusion
by NBP.

In this {\it Letter}, we attempt to settle this longstanding issue based on
the mostly unbiased sample of stars, together with a devised technique
for analyzing the orbital distribution of stars.
Our sample stars, which contain metal-deficient red giants and RR Lyrae
variables in the vicinity of the Sun, are expected to provide
precise chemo-kinematical properties of old stellar populations
compared to those used by ELS and NBP, for the reasons of
their selection without kinematic bias and of
their well-determined metal abundances and distances from recently revised
spectro-photometric calibrations (ATT; Layden 1994; Layden 1996).
Further advantage of using these sample stars is offered by their homogeneity
in the data of proper motions which are accurately measured by
the Hipparcos satellite (ESA 1997), and
their insensitivity to the effects of metal accretion due to
encounters with interstellar clouds because of their deep convection zone
(Yoshii 1981)\footnote{
Whether or not the subdwarf studies (e.g. Carney et al. 1990) are subject to
the effects of metal accretion is yet to be settled.}.
In order to clarify the global orbital properties of these metal-poor stars
in the Galactic space, we adopt, in contrast to most of previous studies,
a realistic, three-dimensional Galactic potential (Sommer-Larsen \& Zhen 1990)
which allows us to establish the three-dimensional orbital motions of stars
in conjunction with their metal abundances.
The extensive analyses and results are reported elsewhere
(Chiba \& Yoshii 1997).

\section{METHOD AND RESULTS}

We have selected 122 red giants from the kinematically unbiased sample
of metal-deficient red giants surveyed by Bond (1980), and 124 RR Lyraes
from the catalogues of variable stars compiled by Kukarkin (1969-1976).
The proper motions of the sample stars have been observed by the Hipparcos
satellite accurate to $\sim 1$ mas/yr, which were allocated to the senior
author's proposal submitted in 1982. The radial velocities and their errors
are assembled from a number of published works. The data on distances
and metallicities are taken from the spectro-photometric studies by ATT for
188 red giants and Bond (1980) for other 4 red giants, and by Layden (1994;
1995) for all RR Lyraes.
The full listing of all data and literatures that we have used will appear
in Chiba \& Yoshii (1997).
A majority of the sample stars are located within about 2 kpc from the Sun and
have metallicities [Fe/H] less than $-1$ dex.
We note that the distribution of their metallicities in the range of
[Fe/H]$\le -1$ is in good agreement with the likely true metallicity
distribution of halo derived from subdwarf stars (Laird et al. 1988).

For the purpose of enlightening the orbital properties of these stars combined
with their metallicities, we calculate the eccentricities $e$ of orbits by
assuming a Galactic potential. We adopt the three-dimensional St\"ackel-type
potential constructed by Sommer-Larsen \& Zhen (1990) which reproduces well
the mass model of Bahcall, Schmidt \& Soneira (1982). Since orbits with
this potential are generally not closed, we arbitrarily define $e$ as
$e = (r_{ap}-r_{pr})/(r_{ap}+r_{pr})$, where $r_{ap}$ and $r_{pr}$ denote the
apogalactic and perigalactic distances, respectively. It is worth noting
that use of other axisymmetric potentials does not modify the essential
results explained below.

In Figure~1(b), we plot the relation between [Fe/H] and $e$
for RR Lyraes (open circles) and red giants (filled circles).
It is apparent that while the stars with [Fe/H]$>-1$ have nearly circular
orbits with $e<0.4-0.5$, those with [Fe/H]$\le -1$ have a diverse range in
orbital eccentricities, i.e. there is no correlation between [Fe/H] and $e$.
Even for [Fe/H]$\le -1.6$, about 16\% of stars have $e<0.4$. The primary
reasons for the change from Figure~1(a) include: (1) sampling the stars almost
completely in accord with the halo's metallicity distribution, especially
those for [Fe/H]$\le -1.6$, whereas the number of such metal-poor stars in NBP
is insufficient,
and (2) adopting a more realistic, three-dimensional Galactic potential
including a massive halo in order to properly take into account their large
vertical velocities above and below the Galactic plane (see Figure~3 below),
whereas $e$ based on a two-dimensional potential adopted by ELS and NBP is
overestimated (Yoshii \& Saio 1979)\footnote{
It is also remarked that NBP selected the stars with probable errors in $e$
less than 0.1. This selection of stars however creates an extra bias in the
sample (Twarog \& Anthony-Twarog 1994; Chiba \& Yoshii 1997).}.
We note that the diverse $e$ distribution for [Fe/H]$\le -1.6$ is not affected
by the systematic underestimation of [Fe/H] for metal-rich stars
with disk-like kinematics (see Twarog \& Anthony-Twarog 1994 for details).

It has been discussed by recent workers that low-$e$, low-[Fe/H] stars
may belong to the metal-weak tail of the rapidly rotating thick disk
which dominates at [Fe/H]$=-0.6$ to $-1$ (NBP;
Beers \& Sommer-Larsen 1995). If this is the case, many
low-$e$ stars in the [Fe/H]-$e$ diagram ought to disappear when we restrict
to the larger height from the Galactic plane than the scale height of the
thick disk with $\sim 1$ kpc (Freeman 1987).
In Figure~1(b), the domain enclosed by dotted lines shows where low-$e$ stars
are selectively excluded at $|z|\ge 1$ kpc.
It is clear that in the intermediate metallicity range of
$-1.6<$[Fe/H]$\le-1$, the thick disk overlaps with the halo at $e \la 0.6$, but
this metal-weak tail of the disk does not extend down to [Fe/H]$\le -1.6$.
We have found that such a division between the halo and thick disk components
at [Fe/H]$\sim -1.6$ is rather sharp (Chiba \& Yoshii 1997).

The above indication that low-$e$ stars in the range of [Fe/H]$\le -1.6$
belong to the halo component is also supported from the following
analysis of orbital distributions. Figure~2(a) and (b) show the cumulative $e$
distributions of orbits $N(<e)$ for [Fe/H]$\le -1.6$ and $-1.6<$[Fe/H]$\le-1$,
respectively. Various histograms show how the distribution changes when we
set limits on the range of the height $|z|$. 
For [Fe/H]$\le -1.6$, the overall $e$ distribution of stars remains essentially
unchanged with increasing $|z|$. Thus the stars in the range of
[Fe/H]$\le -1.6$ belong to the halo component, where a non-negligible fraction
($\sim 16-20$\%) of stars have nearly circular orbits with $e<0.4$.
On the other hand, the number of low-$e$ stars in the range of
$-1.6<$[Fe/H]$\le-1$ shows a sharp drop at $|z|\sim 1$ kpc and then remains
unchanged at higher $|z|$ [Figure~2(b)]. This suggests that the stars with
these intermediate metallicities are contaminated with the thick disk component
with a scale height of $\sim 1$ kpc, except for those stars having high $e$
at high $|z|$. The halo is thus well separated from other disk components if
we set the restriction $|z|\ge 1$ kpc.

Figure~3 further presents the distribution of vertical velocities $V_z$ above
and below the Galactic plane for [Fe/H]$\le -1.6$ (panel $a$) and
$1.6<$[Fe/H]$\le-1$ (panel $b$). Dotted lines correspond to a typical
vertical velocity dispersion ($38$ km s$^{-1}$) of the thick disk
(Beers \& Sommer-Larsen 1995). The figure virtually demonstrates that low-$e$
stars with [Fe/H]$\le -1.6$ have much larger $|V_z|$ than expected from the
thick disk, whereas some fraction of low-$e$ stars with $-1.6<$[Fe/H]$\le-1$
are characterized by low $|V_z|$. These results further strengthen our
conclusion stated above. It is also noted that use of the planar, ELS Galactic
potential to evaluate $e$ is not supported because the orbital motions in the
$z$ direction are closely coupled with those in the plane.

We obtain the fractional value of the thick disk's contribution $F$
in the range of $-1.6<$[Fe/H]$\le-1$. Figure~2(c) indicates that in this
metallicity range, the number of low-$e$ stars at $|z|<1$ kpc is systematically
large compared to that in the range of [Fe/H]$\le -1.6$.
The evaluation of this excess at $|z|<1$ kpc over the $e$ distribution in the
halo-metallicity range of [Fe/H]$\le -1.6$ allows us to obtain $F \sim 0.1$
for $-1.6<$[Fe/H]$\le-1$, and this value is increased to only $F \sim 0.2$
even for $-1.4<$[Fe/H]$\le-1$ (Chiba \& Yoshii 1997). Thus, in contrast to
earlier results suggesting a large $F$ (e.g. $F\sim 0.72$ by Morrison, Flynn
\& Freeman 1990; $F \sim 0.6$ by Beers \& Sommer-Larsen 1995),
the fraction of the metal-weak thick disk is found to be small.

\section{DISCUSSION AND CONCLUSION}

Using the mostly unbiased sample of red giants and RR Lyrae variables,
we arrive at the conclusion that orbital eccentricities of metal-poor
halo stars are no longer correlated with their metallicities.
This result is virtually in contrast to the ELS result,
but is in agreement with other studies using subdwarf stars
(Carney et al. 1990). A detailed analysis of the eccentricity distribution of
orbits, which is made possible for the first time by the accurate Hipparcos
proper motions (ESA 1997) and well-calibrated metal abundances of the almost
complete halo sample (ATT; Layden 1994; Layden 1996), allows us to conclude:
(1) low-$e$ stars in the range of [Fe/H]$\le -1.6$ belong to the halo, and
(2) the metal-weak thick disk constitutes a small fraction down to
its boundary at [Fe/H]$\sim -1.6$. Our sample stars are also less sensitive to
the effects of secondary metal accretion onto the surface of stars
(Yoshii 1981), thereby the currently observed metallicities of these stars can
be regarded to retain the fossil records for those when the stars were formed.

The above conclusions set constraints on the scenarios for the contraction
of the Galaxy at high redshifts. The absence of a correlation between [Fe/H]
and $e$ suggests that the Galaxy has not experienced a free-fall monolithic
collapse (ELS) unless otherwise there remain no metal-poor, high angular
momentum stars after collapse. Alternatively, the Galactic spheroid may
have been assembled from merging or accretion of numerous
fragments (Searle \& Zinn 1978), such as dwarf-type galaxies, but whether or
not the aftermath reproduces the observed kinematical distribution of stars
is yet to be explored (e.g. Steinmetz \& M\"uller 1995).
The early evolution of the Galaxy also involves the stage of the thick-disk
formation. It is also yet unsettled whether or not a leading merger scenario
for forming thick disk (e.g. Quinn et al. 1993) reproduces our finding of
only few stars with disk-like kinematics in the range of [Fe/H]$\le -1$.

Therefore, the comprehensive understanding of the observed chemo-kinematical
properties of metal-poor stars demands
to take more advanced approaches to study the Galaxy formation, perhaps guided
by both extensive numerical simulations of a collapsing galaxy and
kinematical data of more stars in the whole region of the Galactic space
provided by next-generation astrometric telescopes.

\acknowledgments

This work has been supported in part by the Grand-in-Aid for
Scientific Research (08640318, 09640328) and COE Research (07CE2002)
of the Ministry of Education, Science, and Culture in Japan.


\clearpage

\clearpage
\begin{figure}
\plotone{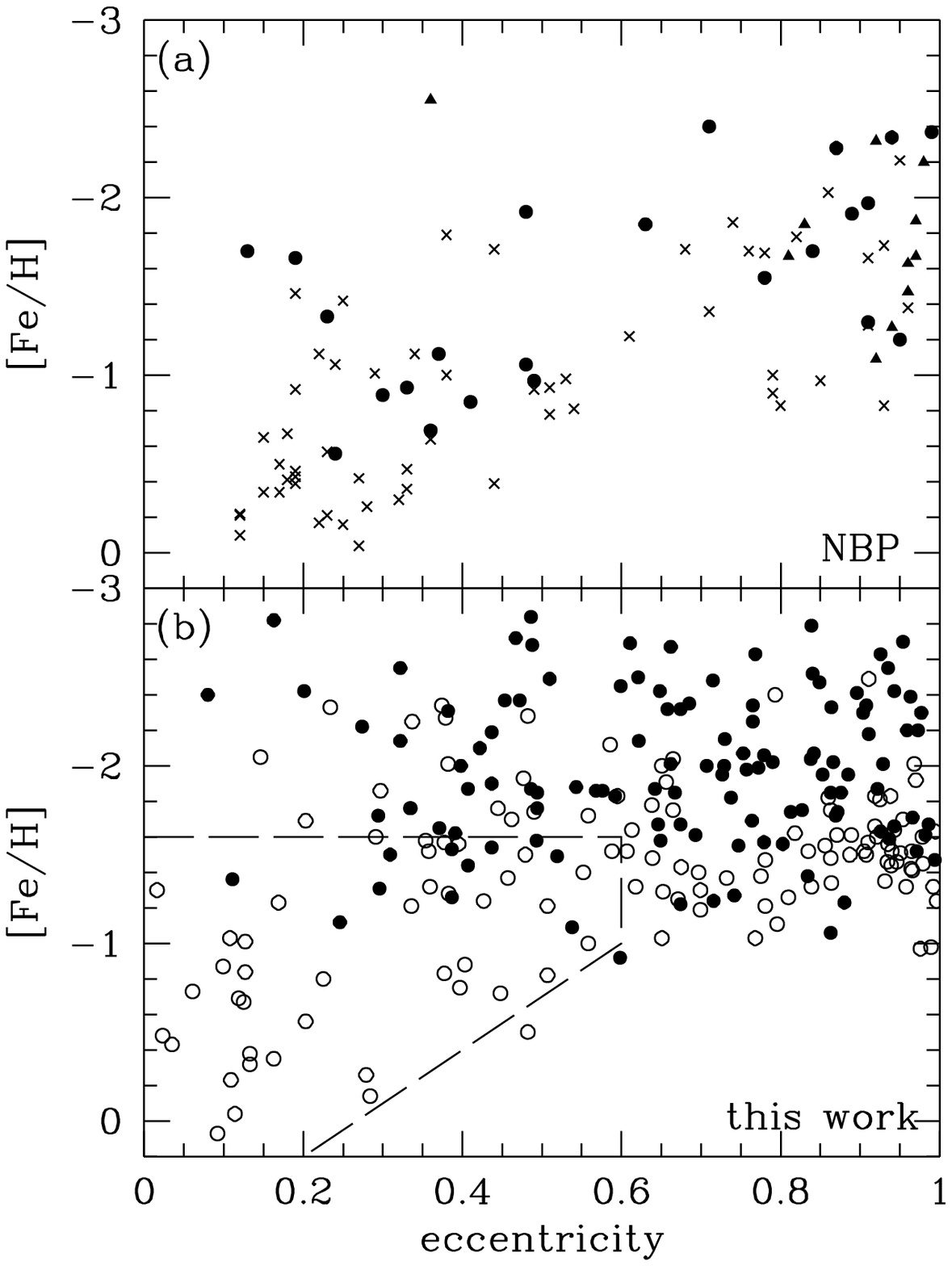}
\caption{
(a) The relation between [Fe/H] and $e$ reproduced from Fig. 14(b) of NBP for
dwarfs (crosses) and red giants (filled symbols). For the latter
stars, we have used the revised metal abundances when they are available
from either ATT or Ryan \& Lambert (1995).
Filled triangles denote the stars which are common to our sample.
(b) The same relation but from our Hipparcos sample for RR Lyrae variables
(open circles) and red giants (filled circles).
The domain enclosed by dotted lines corresponds to that where the stars are
selectively excluded when we set the constraint $|z|\ge1$ kpc. See the text
for details.
}
\end{figure}

\begin{figure}
\plotone{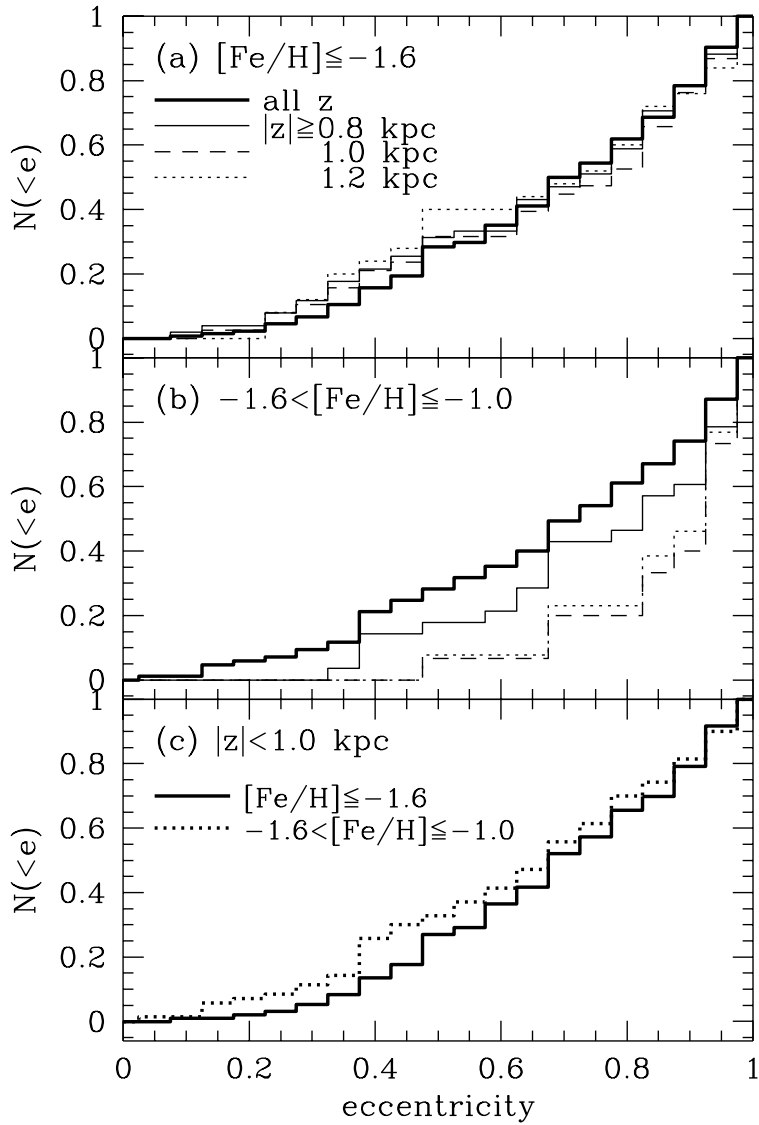}
\caption{
The cumulative $e$ distribution of our sample stars for [Fe/H]$\le -1.6$ (a)
and $-1.6<$[Fe/H]$\le-1$ (b), in the various ranges of the height $|z|$
away from the Galactic plane.
(c) The same distribution when we set the constraint $|z|<1$ kpc, corresponding
to the vertical range of the thick disk component.
}
\end{figure}

\begin{figure}
\plotone{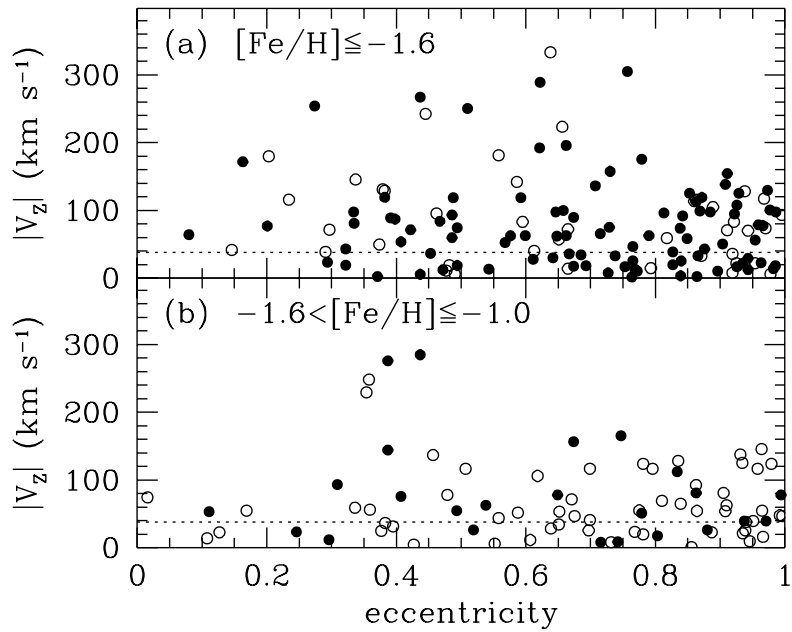}
\caption{
The vertical velocities $|V_z|$ of our sample stars as a function of $e$,
for [Fe/H]$\le -1.6$ (a) and $-1.6<$[Fe/H]$\le-1$ (b).
Dotted lines represent a typical
vertical velocity dispersion of the thick disk
in the $z$-direction.
The symbol designation is the same as in Fig.~1(b).
}
\end{figure}

\end{document}